\documentclass[]{elsarticle}

\usepackage{amsmath}%
\usepackage{amsfonts}%
\usepackage{amssymb}%
\usepackage{float}
\usepackage{amsthm}
\usepackage{graphicx}
\usepackage{float}
\usepackage{color}
\usepackage{hyperref}
\usepackage[normalem]{ulem}
\usepackage{pdfsync}

\usepackage{multicol}
\usepackage{float}

\numberwithin{equation}{section}

\newcommand{\ve}{\varepsilon}
\renewcommand{\epsilon}{\varepsilon}

\makeatletter
\renewcommand{\maketag@@@}[1]{\hbox{\m@th\normalsize\normalfont#1}}%
\makeatother

\makeatletter
\newcommand\footnoteref[1]{\protected@xdef\@thefnmark{\ref{#1}}\@footnotemark}
\makeatother

\begin{document}

\begin{frontmatter}

\title{Dynamics of a cell motility model near the sharp interface limit\tnoteref{t1}} 
\tnotetext[t1]{https://doi.org/10.1016/j.jtbi.2020.110420. This work is licensed under CC BY-NC-ND 4.0. To view a copy of this license, visit https://creativecommons.org/licenses/by-nc-nd/4.0 }
\author{Nicolas Bolle%
\fnref{fn1}}
\ead{bolle.3@osu.edu}

\address{Department of Mathematics and Statistics, The College of New Jersey\\
	Ewing Township, NJ}
	

\author{Matthew S. Mizuhara\corref{cor1}}
\ead{mizuharm@tcnj.edu}
\address{Department of Mathematics and Statistics, The College of New Jersey\\
Ewing Township, NJ}


\cortext[cor1]{Corresponding author}
\fntext[fn1]{Present address: Department of Mathematics, The Ohio State University, Columbus, OH}

\begin{abstract}
	Phase-field models have recently had great success in describing the dynamic morphologies and motility of eukaryotic cells. In this work we investigate the minimal phase-field model introduced in \cite{berlyand2017sharp}. Rigorous analysis of its sharp interface limit dynamics was completed in \cite{MizBerRybZha15,mizuhara2019uniqueness}, where it was observed that persistent cell motion was not stable. In this work we numerically study the pre-limiting phase-field model near the sharp interface limit, to better understand this lack of persistent motion.  We find that immobile, persistent, and rotating states are all exhibited in this minimal model, and investigate the loss of persistent motion in the sharp interface limit.
\end{abstract}

\begin{keyword}
	phase-field \sep keratocyte motion \sep traveling wave \sep rotating cell
\end{keyword}

\end{frontmatter}

%

\section{Introduction}

Eukaryotic cell motility underlies numerous biological processes including the  immune response and cancer metastasis. Cell motion is initiated and maintained by an evolving cytoskeleton comprised of actin and myosin proteins capable of driving a wide range of motility modes. In recent years a variety of modeling techniques have been highly successful in replicating,  explaining, and predicting cell morphologies observed in experimental settings, see, e.g., \cite{aranson2016physical,barnhart2017adhesion,CamZhaLiLevRap16,cao2019cell,Mog09,MogKer09,mogilner2018intracellular,raynaud2016minimal,shao2010computational}. In this work we focus on the bridge between two specific modeling paradigms: free boundary problems and phase-field models. Free boundary problems track the cell boundary via a curve (2D) or surface (3D) whose evolution is governed a geometric evolution equation, often dictated by boundary data of a differential equation solved on the interior. Phase-field models, on the other hand, use an evolving order parameter whose finite width transition layer between phases tracks the cell boundary. Phase-field models avoid difficulties of explicitly discretizing and tracking the moving interface, making them ideal for numerical simulation.

We are motivated by the 2D phase-field model for keratocyte fragments (e.g., lacking a nucleus) studied in \cite{berlyand2017sharp}. It is a minimal version of a more general model introduced by Ziebert, et al. in \cite{Zie12}. The original model in \cite{Zie12} has been extended to include spatial adhesion dynamics \cite{Zie13}, non-homogeneous substrate effects \cite{LobZieAra14,mizuhara2017minimal,reeves2018rotating}, and interacting dynamics of multiple cells \cite{LobZieAra15}. These extended models exhibit a wide variety of dynamical modes and can be used to understand the complex morphologies of dynamics cells. On the other hand, the minimal model of \cite{berlyand2017sharp} allows for rigorous mathematical analyses of the model. For example, in 1D, necessary conditions for the existence and stability of persistent motion were proven \cite{berlyand2017sharp}. {In 2D, sufficient conditions for existence and non-existence of persistent motion were studied in \cite{MizBerRybZha15,mizuhara2019uniqueness}.}

However, a numerical exploration of the simplified phase-field model remains unexplored in 2D. Thus, our primary goal of this work is to numerically study the minimal 2D phase-field model introduced in \cite{berlyand2017sharp}  over a range of parameters. We find that this simplified version is surprisingly capable of exhibiting a range of motions in qualitative agreement with more sophisticated models, including stationary, persistently traveling, and rotating modes. 



The minimal model admits a non-trivial sharp interface limit: an asymptotic reduction of the phase-field model in the limit that the width of the diffuse interface (the location of the cell boundary) tends to zero, transforming the model into a free boundary problem. Rigorous analysis of the sharp interface limit was completed in both 1D \cite{berlyand2017sharp} and in 2D \cite{MizBerRybZha15,mizuhara2019uniqueness}, where sufficient conditions for existence of traveling wave solutions were proved. These analyses were thus able to provide insight into the minimal biophysical mechanisms that are necessary to drive these motility modes, and which modes require more complex mechanisms.

However, previous numerical simulations of the 2D sharp interface limit showed that persistent motion was unstable, and more crucially, does not exist when certain symmetry is present (see Section \ref{sec:conclusion} for details). Therefore, a secondary goal of our work is to explore persistent motion in the phase-field equation, and understand its existence as we approach the sharp interface limit. 

\section{Model}

\subsection{Biological background}
Keratocytes are prototypical for the experimental and mathematical study of cell motion. Their characteristic cell length/width is two orders of magnitude larger  than  the  height  while  motile, hence they are well described by 2D  models (for recent advances in 3D models see, e.g., \cite{Haw11,tjhung2015minimal,winkler2019confinement}). Keratocytes are additionally able to exhibit persistent motion over many times the cell length with approximately constant cell shape, making them ideal starting points for the study of motion \cite{MogKer09,verkhovsky1999self}.   

We recall the following key factors contributing to cell motion, and refer the reader to \cite{Mog09} for a more detailed review. A crawling cell maintains self-propagating motion via internal forces generated by actin polymerization. Actin monomers bind together to form filaments which create a dense network at the leading edge of the cell, known as the lamellipod. The cell's leading edge protrudes via growth of actin laments at the cell membrane and degradation of the filaments towards the interior of the cell, a process known as actin treadmilling. Intercellular adhesion complexes form ligand bonds to the substrate in order to transform this propulsion force into traction forces. Myosin motors interact with actin filaments to generate contractile forces. Acto-myosin interactions contract of the rear part of the cell, pulling the rear of the cell, and allowing for persistent motion. In idealized mathematical settings, such persistent motion is described by traveling wave solutions.  Such motility remains a rich area of study: persistent motion been observed in both myosin-inhibited \cite{herant2010form}, and actin-inhibited \cite{hirsch2017mathematical,RecPutTru15} cells. It has additionally been observed that random fluctuations are sufficient to spontaneously switch a cell from a symmetric, non-motile state to motile states \cite{BarLeeAllTheMog15,Haw11}.

Moreover, by varying biophysical parameters (such as actin polymerization strength, substrate adhesion/elasticity properties, or myosin motor strength), cells additionally exhibit a wide range of motility modes beyond persistent motion, such as stick-slip (oscillations in translational velocities) and bipedal (left and right sides alternating forward motion) motions \cite{BarAllJulThe10,Bar11}. As such, there is a deep need to understand the interactions of various biophysical pathways leading to such a variety of behaviors.

Finally, of particular interest are ``rotating'' cells, experimentally observed in \cite{Lou15}, where
cells remained essentially stationary but experienced lateral waves of protrusions of the membrane.
This was caused by the expression of a particular kinase (MLCK) leading to an increase of myosin activity in the cell's lamellipod. {The increased myosin activity was hypothesized to enhance actin depolymerization (or alternatively, to sweep the actin network toward the center of the cell), thus resulting in shorter edge protrusion lifetimes. Shorter edge lifetimes cannot sufficiently polarize the cell to generate motion in a single direction, resulting in lateral waves of actin protrusion at the cell boundary.}

\subsection{Phase-field model}
{
The following 2D phase-field model for cell motility was originally introduced in \cite{Zie12}; we study a slightly simplified form (e.g., omitting some non-linearities) which is amenable to rigorous analysis (see Section \ref{sec:sil}):
\begin{align}\label{original1}
\partial_t \rho &= D_\rho \Delta \rho - \tau_1^{-1}F(\rho) -\alpha (\nabla \rho)\cdot {\bf P}\\
\label{original2}\partial_t {\bf P} &= D_P \Delta {\bf P} -\tau_2^{-1} {\bf P} -\zeta\nabla \rho.
\end{align}
Here
\begin{align}
F(\rho) &= (1-\rho)(\delta(\rho)-\rho)\rho,\\
\delta(\rho)&= \frac{1}{2}+\mu\left[ \int_\Omega (\rho(x,y,t)-\rho(x,y,0))\; dxdy \right].
\end{align}
The phase-field variable $\rho=\rho(x,y,t)$ takes value $\rho\approx 1$ in the interior of the cell and $\rho\approx 0$ outside the cell. The area of the cell,
	\begin{equation}
	A(t) = \int_\Omega \rho(x,y,t) dxdy,
	\end{equation} 
	is approximately preserved due to penalization from $\delta$, see \cite{Zie12} for more details. Descriptions of parameters and typical values for keratocytes are presented in Table \ref{tab:1}. 
}

	\begin{table}
		\centering
		\begin{tabular}{lll}
			Parameter & Description & Typical values \\
			\hline 
			$D_\rho$ & interface stiffness & 1 $\mu m^2/s$\\
			$\tau_1$ & time scale of curvature motion & 1 $s$ \\
			$\alpha$ & adhesion site formation rate	& .5 - 3 $\mu m/s$	\\
			$D_P$ & actin diffusion & $.1$ $\mu m^2/s$ \\
			$\tau_2 $ & actin depolymerization & $10$  $s$ \\
			$\zeta$ & actin polymerization rate & $1$ - $2$  $\mu m/s$\\
			$R$ & cell size (radius) & $5$ - $18$ $\mu m$
		\end{tabular}
	\caption{Parameter descriptions of original phase-field model \eqref{original1}-\eqref{original2} together with typical values, from \cite{Zie12}.}
	\label{tab:1}
	\end{table}

{
  The interface stiffness $D_\rho$ represents the ratio of surface tension of the cell membrane to friction with the substrate \cite{shao2010computational,Zie12}. Together $D_\rho$ and $\tau_1$ dictate the rate of passive curvature driven motion and the size of the diffuse interface. In particular the width of the diffuse interface scales with $\sqrt{D_\rho \tau_1}$ \cite{ziebert2016computational}. Active transport of $\rho$ occurs along the vector field ${\bf P}={\bf P}(x,y,t)$ which represents {the actin network from a macroscopic point of view. At each $(x,y)$ the vector field points in the average local orientation of actin filaments, and the magnitude of the vector represents the degree of orientation (e.g., many well-aligned actin filaments result in a larger magnitude vector)} \cite{Zie12}. 
}

Physically, advection requires formation of substrate adhesions to generate traction forces. {To that end, a transmembrane complex of proteins form ligand bonds to the substrate and connect to the actin network.} In \cite{Zie13} adhesion dynamics are explicitly modeled via a PDE for an auxiliary scalar field $A=A(x,y,t)$ whose dynamics additionally encode substrate deformations (as a visco-elastic medium), and transport requires adhesion formation so that $\partial_t \rho \sim  A {\bf P}\cdot \nabla \rho$. For simplicity, we assume that adhesion is formed instantaneously and uniformly with the vector field ${\bf P}$. 

Dynamics of actin filaments ${\bf P}$ are regularized by diffusion and experience global decay due to depolymerization. As actin polymerization is localized to the boundary of the cell, the source term for ${\bf P}$ is given by $-\zeta \nabla \rho$, where $\zeta$ is the rate of actin polymerization.


Competition between advection by ${\bf P}$ and curvature motion flow from the Allen-Cahn contribution constitute the main dynamics of interest: one expects that if $|{\bf P}|$ is sufficiently small then the cell remains immobile and if $|{\bf P}|$ is sufficiently large then the cell has sufficiently many active internal forces to generate motion. 

In \cite{reeves2018rotating}, numerical simulations of the more general phase-field model, including additional non-linear effects from heterogeneous myosin contraction, non-linear dynamics of adhesion complex formation, and substrate viscoelasticity. In that more complex setting, they observe several motility modes including several types of rotating lamellipod solutions.

\subsection{Non-dimensional model} {To study the sharp interface limit, one first non-dimensionalizes and chooses an appropriate scaling to give rise to a non-trivial limit. To that end introduce parameters
\begin{equation}\label{non-dim}
	\ve := \frac{\sqrt{D_\rho \tau_1}}{R},\;\; \tilde{\alpha} := \frac{\alpha R}{D_\rho},\;\; \tilde{\zeta} := \frac{\zeta \tau_1}{R},\;\; \beta := \frac{\tilde{\alpha}\tilde{\zeta}}{\ve^2},
\end{equation}
where $R$ is the characteristic size of the cell (say the radius). Then, under certain parameter assumptions, \eqref{original1}-\eqref{original2} can be  written in non-dimensional form \cite{berlyand2017sharp}:
\begin{align}\label{eq1}
\partial_t \rho &= \Delta \rho -\frac{1}{\ve^2} W'(\rho) - {\bf P}\cdot \nabla \rho +\lambda\\
\partial_t {\bf P} &= \ve \Delta {\bf P} - \frac{1}{\ve} {\bf P} - \beta \nabla \rho, \label{eq2}.
\end{align}
See \ref{appendix} for a detailed derivation. Using typical values of parameters from Table \ref{tab:1} we see $\ve \sim .1$ and $\beta \sim 100$ for keratocytes.
}

{
 The non-dimensional parameter $\ve$ represents the relative size of the diffuse interface width (to the cell size). As actin polymerization occurs only where $|\nabla \rho|\neq 0$, the parameter $\ve$ also then dictates the width of the region where polymerization occurs. While one would ideally like to decouple surface tension from actin polymerization, this would require complications of the model beyond the scope of this work. The non-dimensionalized polymerization $\beta$ is a ratio of non-dimensional adhesion site formation and actin polymerization rates to $\ve^2$. In original parameters $\beta= \frac{\alpha \zeta R^2}{D_\rho^2}$ and thus it can be thought of as analogous to a P\'eclet number, with advection driven by actin polymerization.}

{
 In \eqref{eq1}-\eqref{eq2} the phase-field parameter $\rho$ now has an $O(\ve)$ thick transition layer and the sharp interface limit can be analyzed as $\ve \to 0$. Since analysis in the sharp interface limit assumes constant $\beta$, in the sharp interface limit we require that the product $\tilde{\alpha}\tilde{\zeta}\to 0$. For example, in the limit of vanishing cell surface tension we have $D_\rho \to 0$ (and thus $\ve\to 0$). Then, one may scale $|\zeta|\sim |D_\rho^2|$ to ensure $\beta$ is kept constant. Naively, one may expect that if $\zeta\to 0$ or $\alpha\to 0$ that no motion would be possible in the sharp interface limit, however, rigorous analysis has revealed that traveling wave solutions exist provided the ratio $\beta$ is sufficiently large (see Section \ref{sec:sil}).
}

{We remark that for simplicity of analysis, area preservation in \eqref{eq1}-\eqref{eq2} is enforced via the Lagrange multiplier
	\begin{equation}
	\lambda(t) = \frac{1}{|\Omega|} \int_\Omega \frac{1}{\ve^2}W'(\rho) +{\bf P}\cdot \nabla \rho \;dx,
	\end{equation}
	so that $W$ is fixed to be the standard Allen-Cahn double-well potential $W(z) = \frac{1}{2} z^2(1-z)^2$. 
	We additionally note that $\lambda = \lambda(t)$ passively encodes myosin motor effects, by virtue of enhancing contraction of the cell membrane (e.g., when $\lambda<0$). Of course, this is vastly simplified as it is assumed to be constant in space and varying only in time.}

\begin{figure}[h]
	\centering
\includegraphics[width = \textwidth]{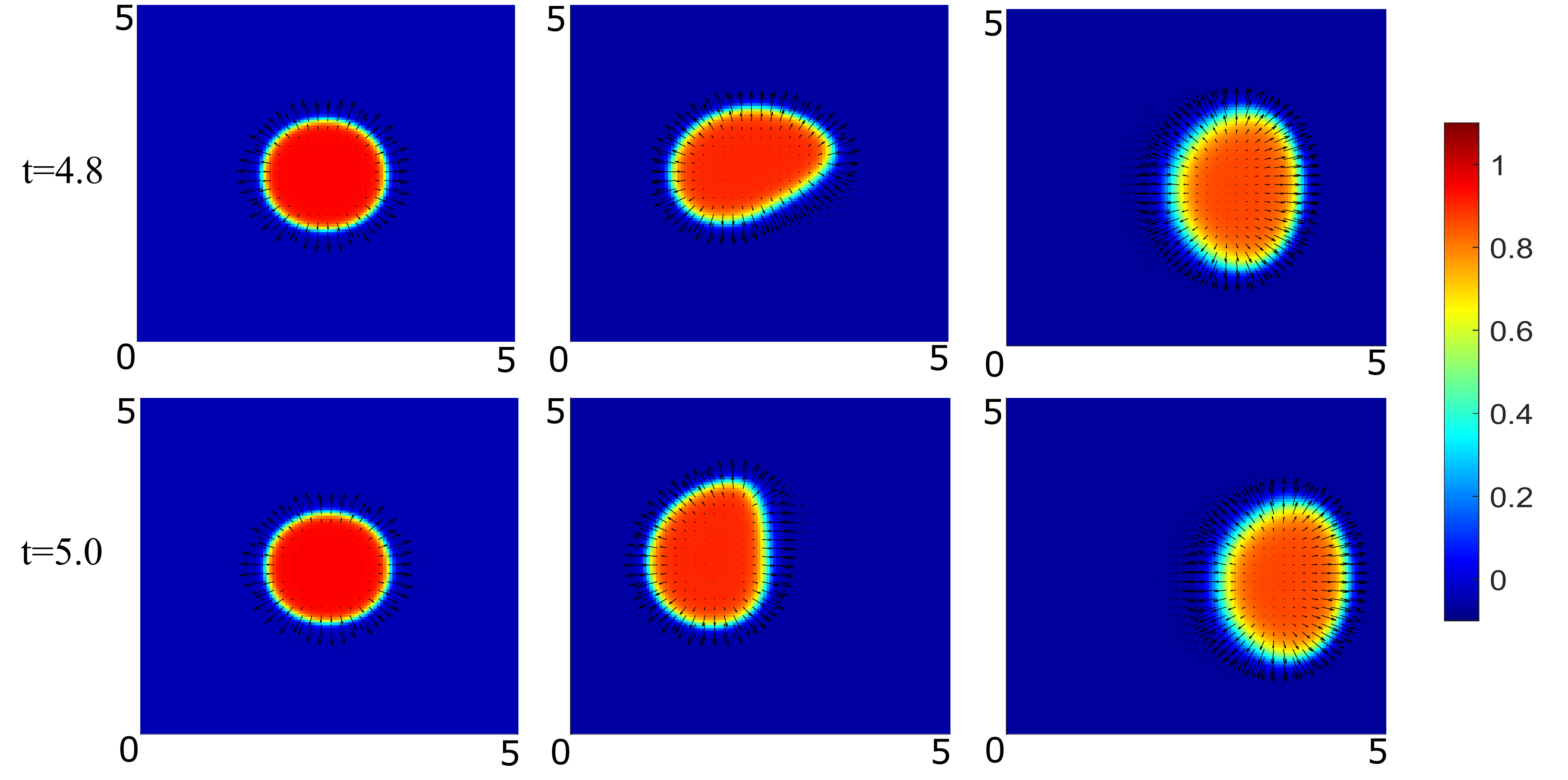}
	\caption{Sample snapshots of the three long time behaviors arising from \eqref{eq1}-\eqref{eq2} at two nearby times: (left) stationary ($\ve=.02$, $\beta = 80$), (center) rotating ($\ve=.035$, $\beta = 110$), and (right) persistent motion ($\ve=.05$, $\beta = 130$). Color indicates value of $\rho$ and arrows represent the vector field ${\bf P}$.}
	\label{fig:tw}
\end{figure}

\subsection{Sharp interface limit}\label{sec:sil}

One reason we study the system \eqref{eq1}-\eqref{eq2}, rather than the more general model, e.g. in \cite{reeves2018rotating}, is that the system \eqref{eq1}-\eqref{eq2} is amenable to rigorous mathematical analysis: it is possible to derive dynamics in the limit $\ve \to 0$ recovering the so-called {\em sharp interface limit}. In \cite{berlyand2017sharp}, both the pre-limiting and sharp interface limit dynamics in 1D are rigorously studied. In particular necessary conditions for the existence and stability of traveling wave solutions are established, corresponding to persistently traveling cells. Additionally, Berlyand et al. establish the 2D sharp interface limiting equation, recovering a geometric evolution equation for planar curves, $\Gamma(s,t)$ representing the boundary of the cell. They show that the normal velocity $V$ of the curves at each moment evolve by
\begin{equation}\label{eq:sil}
	V = \kappa + \beta\Phi(V) -\frac{1}{|\Gamma(t)|}\int_\Gamma \kappa + \beta\Phi(V) ds,
\end{equation}
where $\kappa=\kappa(s,t)$ represents the curvature at location $s$ and time $t$, and $\Phi\colon \mathbb{R}\to\mathbb{R}$ is a fixed non-linear function whose form is explicit and depends on the double-well potential $W$. The integral term, as before, enforces volume preservation. Analysis and numerical simulation of \eqref{eq:sil} was completed in \cite{MizBerRybZha15,mizuhara2019uniqueness}; we briefly review the relevant results.

Due to the importance of persistently moving cells, the authors considered traveling wave solutions of \eqref{eq:sil}. These are solutions of the form
\begin{equation*}
	\Gamma(s,t) = \Gamma_0(s)+{\bf v}t,
\end{equation*}
where ${\bf v}$ is a fixed vector representing the velocity of the cell and $\Gamma_0(s)$ is the unknown cell shape. As expected by physical considerations, we require sufficiently large $\beta$ in order for traveling wave solutions to exist, as $\beta$ captures the {relative strength of} actin polymerization. Surprisingly, it was also shown that if $\Phi$ was an even function then traveling waves could not exist, regardless of the value of $\beta$. In particular, the standard Allen-Cahn potential, as considered in the current work, results in even $\Phi$. 



It is thus natural to suspect that the simplified phase-field model \eqref{eq1}-\eqref{eq2} cannot support persistent motion, perhaps because we have eliminated symmetry breaking effects of myosin motors. However, as we will find below, the minimal phase-field model is capable of exhibiting a wide range of motions for various parameter regimes, thanks to the finite width transition layer allowing for non-trivial dynamics of the vector field ${\bf P}$.

\section{Numerical simulation of the phase-field model near the sharp interface}


To understand dynamics near the sharp interface limit, we investigate the long-time dynamics of the phase-field model for various values of $\beta$ and for small values of $\varepsilon$.  {Simulations of} \eqref{eq1}-\eqref{eq2} {are done using an explicit finite difference method with centered space steps ($h=.04$). The domain was the square $\Omega = [0,5]^2$ with periodic boundary conditions. Time steps ($dt =$ 8e-5) are taken sufficiently small to ensure convergence of the simulations: taking smaller time steps did not qualitatively affect any results. Additionally  the cell area is small compared to the domain size to ensure that there were no boundary effects caused by the periodic boundary conditions. Again, taking larger domain size does not qualitatively change dynamics. The non-local term is approximated by Riemann sums (left or right hand choice was irrelevant due to simulations being on a torus). We numerically tracked total enclosed area, $ A(t)=\iint \rho(x,y,t)dxdy$, to ensure that it was conserved over time. {As $\ve$ varies, while the diffuse interface width varies, we emphasize that the enclosed area does not change, as $A(t)\approx A(0)$ is determined by initial conditions alone}. The total time of integration is $T=5$. 
}

Initial conditions for the phase-field are a circular cell. We consider both polarized and non-polarized initial conditions for the actin field: for polarized initial conditions we assume that the actin field on the interior of the cell is constant and pointing to the right, with a small random perturbation to ensure robustness of the results. For non-polarized initial conditions we take random initial conditions for the actin field with $|{\bf P}|\ll 1$. Moreover, we assume sufficiently long time integration to ignore transient effects. We find that, regardless of initial conditions, the long time behavior was qualitatively the same. Thus, we report only on the results of the polarized initial conditions.

\begin{figure}[h]
	\centering
\includegraphics[width=\textwidth]{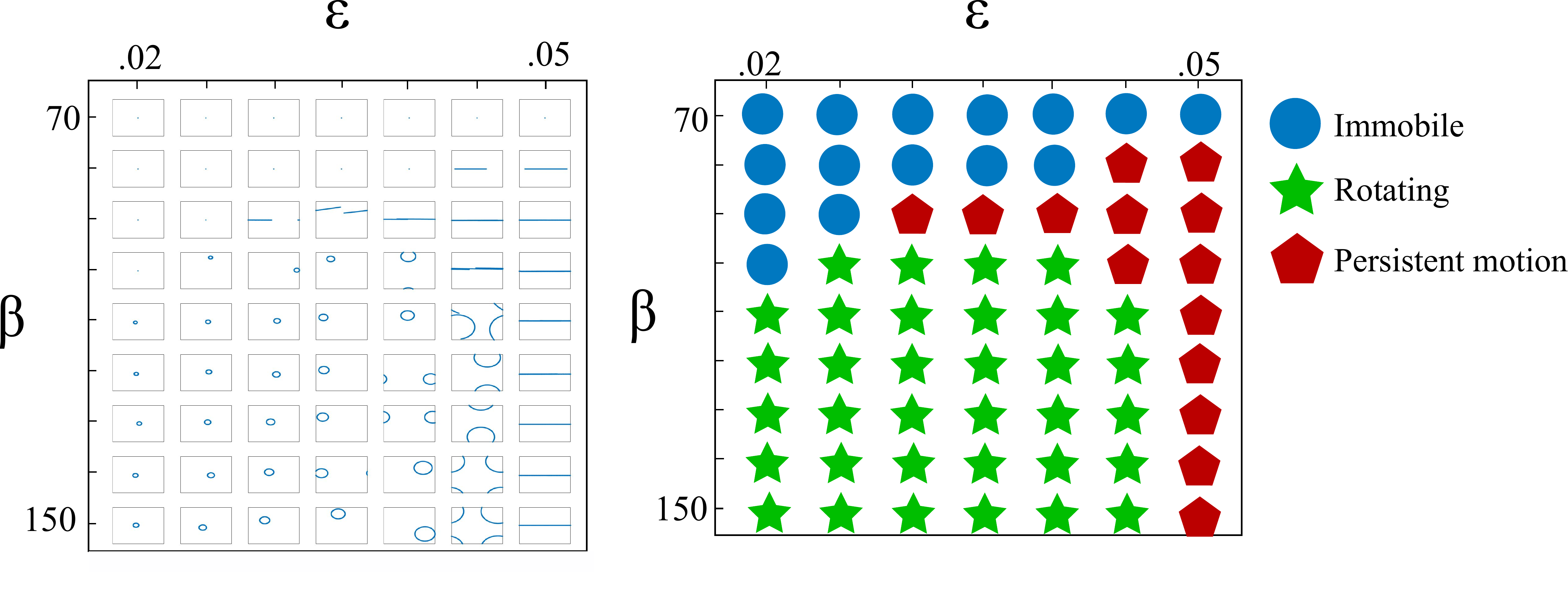}
\caption{Long time behavior of center of mass for a range of parameters. We observe persistent motion (cells moving with constant shape in a straight line), stationary states (cells relaxing to a circular shape without any motion), and rotating motion (asymmetric cells whose center of mass traces a circular path). (left) Trajectories of the center of mass after a transient period and (right) classification of type of motion.}
\label{fig:summary}
\end{figure}

After integrating past a transient time for the cell to reach stable behavior, we record the resultant dynamics. The resultant data is summarized in Figure \ref{fig:summary}. We note that taking smaller values of $\varepsilon$ became computationally expensive, due to the singular nature of the evolution of \eqref{eq1}-\eqref{eq2}.


For all solutions which had non-trivial motion, we track the center of mass of the cell in order to study its trajectory. We then calculate the average radius of curvature of this trajectory in order to distinguish between persistently moving cells (straight line trajectory) and rotating cells (circular trajectory). There is an arbitrary distinction between rotating and persistently moving cell, as a persistently moving cell may have a slight turn due to numerical artifact or non-symmetric initial conditions. Thus, to distinguish the two cases we set a threshold radius of curvature to be $10\sqrt{|\Omega|}$: any trajectories with larger radii of curvature are defined to be persistently traveling.

Over the parameter range considered, we observe three modes of motion: {\bf immobile, persistent, and rotating solutions}. Figure \ref{fig:tw} presents a snapshot and parameter values giving rise to each type of motion.

{\bf Immobile.} We find that for any fixed $\ve$, there is a critical $\beta_{cr}(\ve)>0$, so that for any $\beta<\beta_{cr}$ no motion is possible. This agrees qualitatively with the theory developed in the sharp interface limit, as discussed in Section \ref{sec:sil}. Heuristically, stationary states are expected for sufficiently small $\beta$, since taking $\beta=0$, the model simplifies to the volume preserving Allen-Cahn equation (as ${\bf P}=0$) which asymptotically relaxes to circular steady states. Then, small values of $\beta$ constitute a regular perturbation and so stationary states are expected to persist for sufficiently small $\beta$.

{\bf Persistent motion.}  Persistent motion represents a cell traveling with constant shape. 
Since we use a symmetric double-well potential $W$, it is perhaps surprising, given the theory developed in the sharp interface limit, that any persistent motion is possible. We do observe that for sufficiently small values of $\ve$ these persistent motions no longer exist in our model, agreeing qualitatively with the results of \cite{mizuhara2019uniqueness}. We  conclude that stabilization of persistent motion in the sharp interface limit requires additional myosin motor effects \cite{wilson1991increase}, {or perhaps alternative scalings to be considered in \eqref{eq1}-\eqref{eq2}.}

{\bf Rotating states.} Rotating states indicate that the effective actin polymerization strength is not sufficiently strong to overcome surface tension, creating a lateral wave of actin propagation along the cell boundary and resulting in a rotating wave of protrusion in the cell, which results in a rotating solution.
{Simulations show that rotating states emerge when $\ve$ is relatively small. Since $\ve$ scales with surface tension and $\tilde{\alpha}\tilde{\zeta} = \beta\ve^2$, we see that when $\ve$ is relatively small we have, e.g., $\tilde{\zeta} = O(\ve^2)$ so that effective actin polymerization is a further order of magnitude smaller.}
Traveling actin waves leading to such rotating states have been investigated in the more complex version of the phase-field model in \cite{reeves2018rotating}. 
Indeed, in \cite{reeves2018rotating}, the formation of such waves was explained via shockwaves from a Burger-like equation, whose non-linear shocks are driven by a quadratic contribution $\sim |\nabla \rho|^2$.

{Moreover, rotating cells in experiments were explained via a shortening of edge protrusion lifetime }\cite{Lou15}. {In numerical simulations, since protrusions are generated exclusively from the vector field ${\bf P}$, we assume that the edge protrusion lifetime is related to the time-scale of non-trivial dynamics of ${\bf P}$. To that end, note that on the interface $|\nabla \rho| = O(1/\ve)$, so $\ve |\nabla \rho| = O(1)$. Thus freezing $\nabla \rho$ for simplicity and writing $ \partial_t {\bf P} =\frac{1}{\ve}(\ve^2 \Delta {\bf P} - {\bf P} -\beta \ve \nabla \rho)$, we see that the time-scale of ${\bf P}$ relaxing to equilibrium is also $O(\ve)$. Thus small values of $\ve$ correspond to short edge protrusion lifetimes. 
	{Again, we emphasize} that $\ve$ {dictates} both the time scale of dynamics of ${\bf P}$ as well as the width of the interface layer. As aforementioned, the particular scalings of $\ve$ in the model are required for comparison to the sharp interface limit. To better understand and explain the onset of rotational motion in the phase-field model, one must relax these coefficient restrictions to decouple these two effects. This is beyond the scope of the present work (whose focus is the relationship to the sharp interface limit), though this is an important question for future work.
}

We remark that we did not observe more complicated morphologies such as ameobid or two wave solutions with both actin waves traveling in the same direction, as observed in \cite{reeves2018rotating}. This suggests that such dynamics are driven by more complicated biophysical mechanisms, including hetereogeneities in the myosin motor density or in the adhesion complex formation. 
However, it is surprising that rotating states exist in our simplified model. Our work thus suggests that even with severely weakened myosin contraction, such motility may already be possible in cells.

\subsection{Monotone dependence of trajectory curvature in the sharp interface limit}

While we have already established that persistent motion does not seem possible for any fixed $\beta$ as $\ve \to 0$, we further analyze the trajectory data of simulation results as $\ve \to 0$.

\begin{figure}[h]
	\includegraphics[width = .47\textwidth]{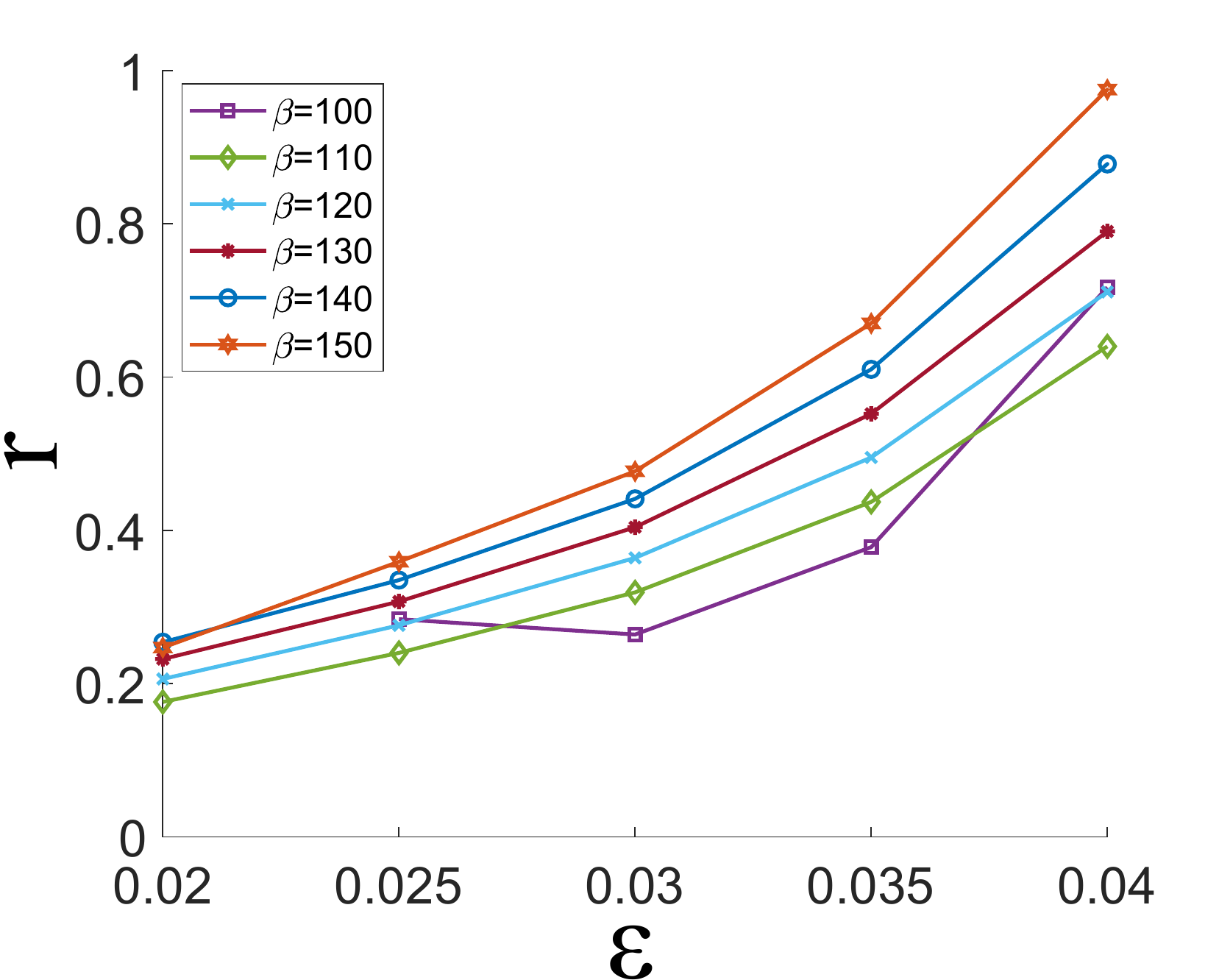}
	\includegraphics[width=.47\textwidth]{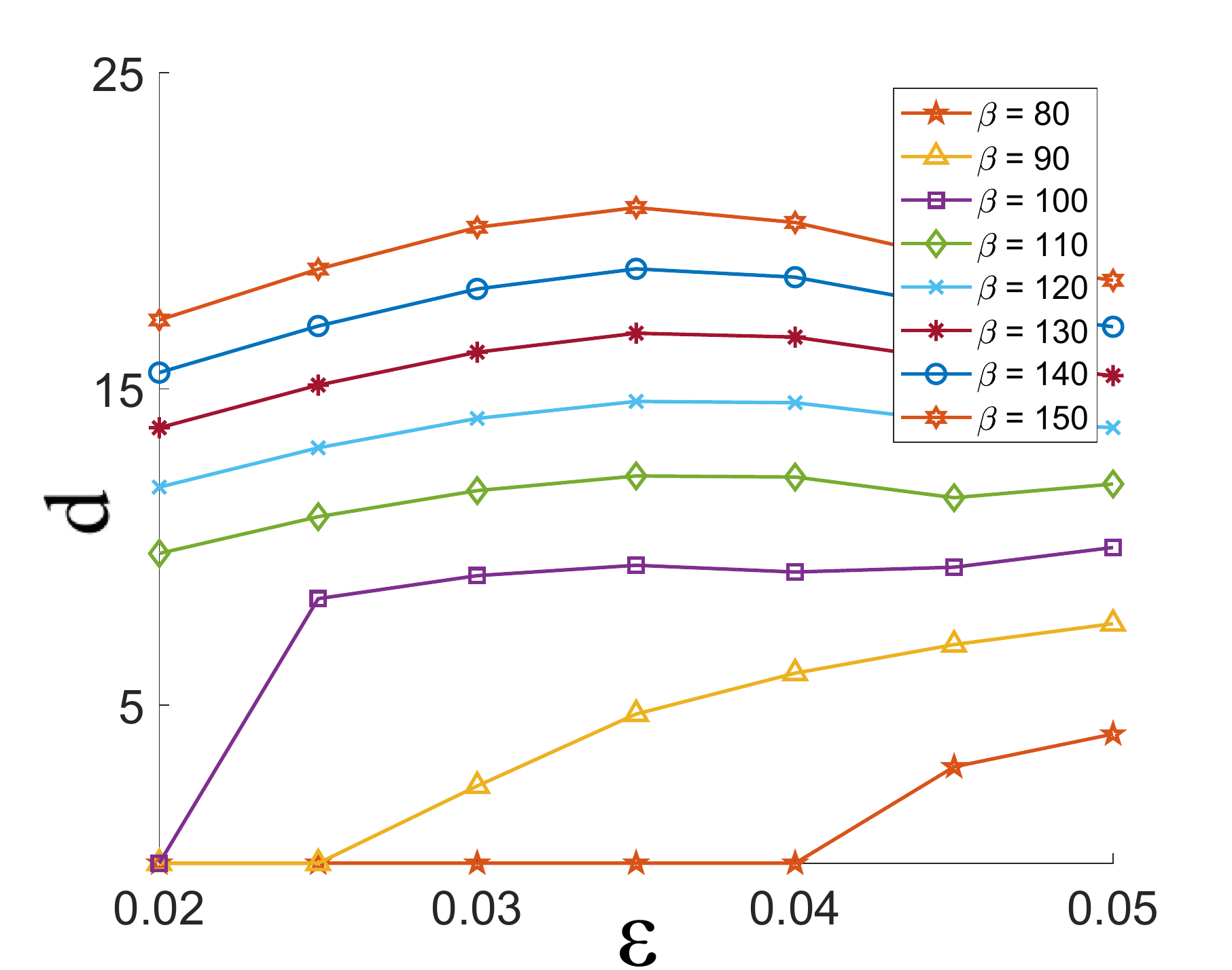}
	\caption{(left) Average radius of circular trajectory for rotating solutions. The data show that, for fixed $\beta$, decreasing $\ve$ results in a smaller radius of curvature. That is, rotating cells trace smaller circles. This suggests that for fixed $\beta$, as $\ve\to 0$, the cell becomes stationary. (right) Distance traveled for both rotating and traveling states show non-monotone behavior as $\ve$ is varied: at $\ve = .035$ we observe (for sufficiently large $\beta$) that distance traveled is maximized. On the other hand, for fixed $\ve$ dependence on $\beta$ is monotone.}
	\label{fig:data}
\end{figure}

To that end, for cells classified as rotating we calculate the radius of curvature of the center of mass's trajectory. Our data reveals that decreasing either $\ve$ or $\beta$ results in a decrease in the radius of the curvature of the trajectory, see Figure \ref{fig:data}. This suggests that as $\ve\to 0$, the cell has a stationary center of mass. While certainly this does not preclude the existence of other motile cells with stationary center of mass, we did not observe any such modes in our simulations.

These data in particular provide evidence for why the sharp interface limit \eqref{eq:sil} may not exhibit persistent traveling waves: for small $\ve$ we see that the only stable states which seem to survive are stationary states and rotating states. As $\ve\to 0$, even rotating states have center of mass trajectories with smaller and smaller radii. So, in the limit $\ve\to 0$ one expects that only stationary states to remain stable. Thus, to exhibit stable persistent traveling wave solutions in the sharp interface limit, one must include other biophysical mechanisms into the analysis of the sharp interface limit.


\subsection{Non-monotone dependence of distance traveled on parameters}

\begin{figure}[h]
	\centering 
	\includegraphics[width = .47\textwidth]{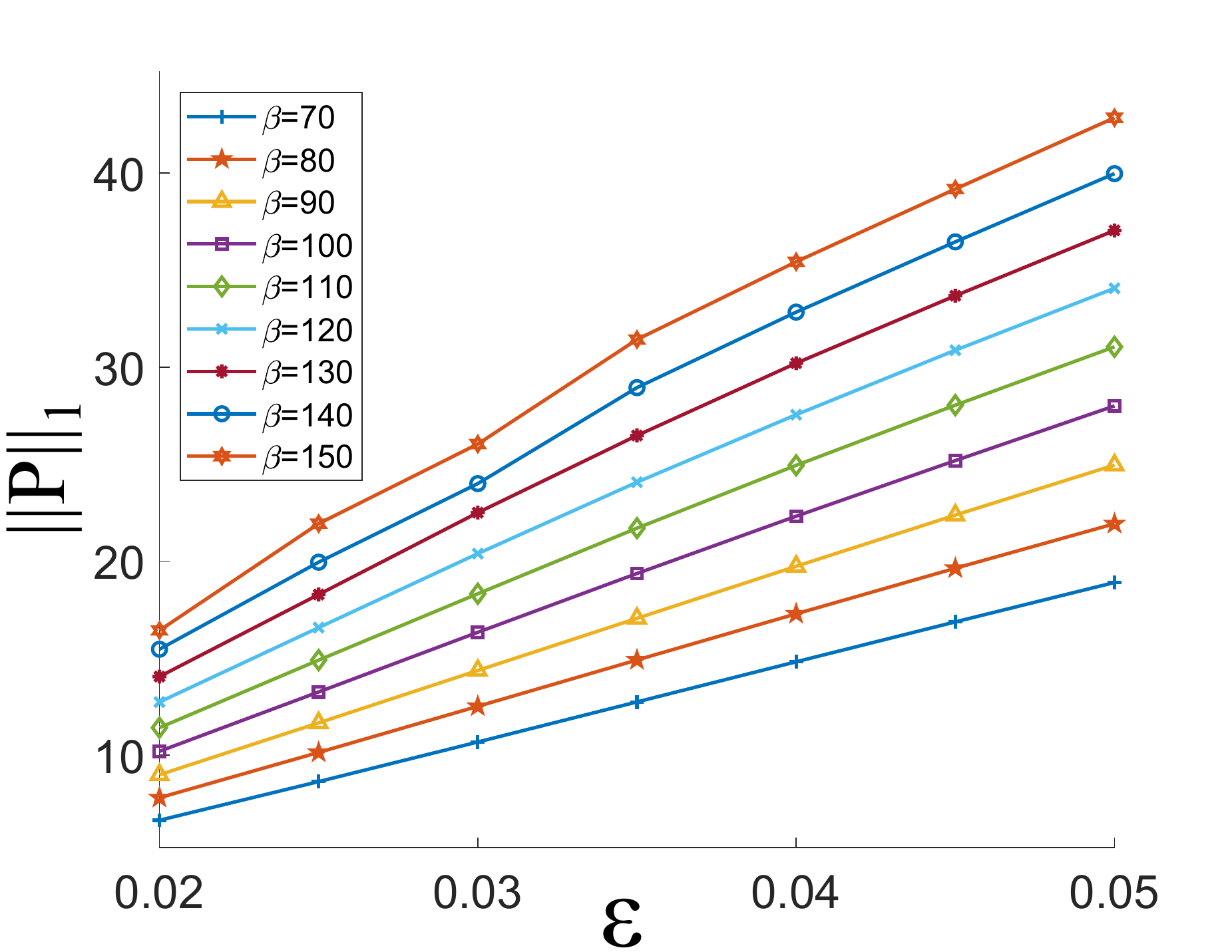}
	\caption{The $L1$ norm of ${\bf P}$ at steady state (after transient effects) shows that as $\ve$ increases, the total amount of actin increases. We interpret this as larger lamellipod size resulting in a larger region of actin polymerization and thus stronger protrusion forces.}
	\label{fig:integralP}
\end{figure}

We investigate the dependence of cell speed on model parameters. To that end, we integrate \eqref{eq1}-\eqref{eq2} to the end time $T=5$ and omit the first $75\%$ of data (to disregard transient effects). Then calculate the distance traveled by the center of mass's trajectory, ${\bf s}(t)$:
\begin{equation*}
	d = \sum |{\bf s}(t_{i+1})-{\bf s}(t_i))|,
\end{equation*}
see Figure \ref{fig:data}.
 We observe that the relationship between distance traveled and parameters is dependent on the motility mode: over parameter ranges where the cell is traveling persistently, if one fixes $\beta$ and increases $\ve$ then the distance traveled  increases monotonically. 
  {Indeed, at the conclusion of each simulation, we also compute}
\begin{equation*}
	\|{\bf P}\|_{1} := \iint |{\bf P}(x,y)| dxdy.
\end{equation*}
{This quantity can be thought of as a measure of {the total forces from actin polymerization in the cell}. We indeed found that for fixed $\beta$, an increase in $\ve$ resulted in an increase in $\|{\bf P}\|_1$, see Figure }\ref{fig:integralP}.

However, over parameter ranges where the cell is rotating there is non-monotone dependence on the distance traveled. In that case, the distance traveled is maximized when $\ve = .035$, provided $\beta$ is large enough. {As $\ve$ is the ratio of diffuse interface width (itself a ratio of membrane surface tension to substrate friction) to cell size, this suggests that a ratio of the cell size to the cell surface tension may be an optimizable quantity for maximizing cell speed. Since our numerics suggest that the optimal $\ve$ is constant for large enough $\beta$, we conjecture that this optimal ratio may be independent from actin polymerization and adhesion site formation rates (provided they are sufficiently large). To our knowledge, this has not been explored experimentally, and thus merits study.}

\section{Conclusion}\label{sec:conclusion}

 {In this work we numerically investigate the  phase-field system} \eqref{eq1}-\eqref{eq2}, {with particular interest in the dynamics near the sharp interface limit $\ve\to 0$. {On the one hand, this explores the possible behaviors of cells upon variation of biophysical parameters. On the other hand this limit is of pure mathematical interest.} The sharp interface limit dynamics} \eqref{eq:sil} {has been studied numerically and analytically in previous works }\cite{MizBerRybZha15,mizuhara2019uniqueness}. {Previous simulation of the sharp interface limit itself required a so-called ``intermediate'' system which is between the full phase-field model and the sharp interface limit (see }\cite{mizuhara2019uniqueness}). {The intermediate system evolves a discretized planar curve, whose evolution requires solving a singularly perturbed parabolic PDE at each discretization point. The system contains a parameter $\delta>0$ (originally $\ve$, but we change notation here for clarity), whose limit as $\delta\to 0$ formally results in the sharp interface limit equation; in that context we note that the value $\delta>0$ is not the width of the transition layer, but was introduced to resolve non-uniqueness of solutions in the sharp interface limit equation. Simulations of the intermediate model showed that if $\beta<\beta_{cr}$ then all dynamics relaxed to circular states. For $\beta>\beta_{cr}$ if traveling wave solutions existed they were unstable. Instead, all dynamics resulted in one of two long term dynamics. For larger values of $\delta$, simulations gave rise to cells moving in a bipedal fashion} \cite{Bar11}. {For smaller values of $\delta$, the intermediate system gave rise to rotating cells. Moreover the rotating cells became more circular as $\delta\to 0$. Importantly, all non-trivial motions arose only when considering asymmetric double-well potentials.}
 
 {Our numerics agree qualitatively with previous results of the intermediate system in many ways. We find that for sufficiently small $\beta$ all cells converge to circular stationary states. We also observe rotating cells which become closer to circular near the sharp interface limit, providing further evidence that the sharp interface limit cannot support non-trivial motion. On the other hand there are some significant differences in the observed dynamics: the current phase-field model does not present bipedal cells, but we do observe persistently traveling cells which are not possible in the intermediate system. Most importantly, traveling waves are possible in the phase-field model with symmetric double-well potential $W$. Analysis of} \eqref{eq:sil}{shows that symmetric potential $W$ (as in} \eqref{eq1}) {cannot give rise to traveling wave solutions in the sharp interface limit, and they did not appear in the intermediate system at all. Moreover, as this system is a minimal version of the work in }\cite{Zie12} {there was no evidence that it could support persistently traveling solutions. Thus their existence in the current model is particularly surprising. A more careful quantitative comparison of the intermediate system with the phase-field model would be of particular interest. However, current numerical methods to solve the intermediate system require weeks to run for a single parameter set, and as such direct comparisons remain unfeasible.}

\begin{figure}[h]
\includegraphics[width = \textwidth]{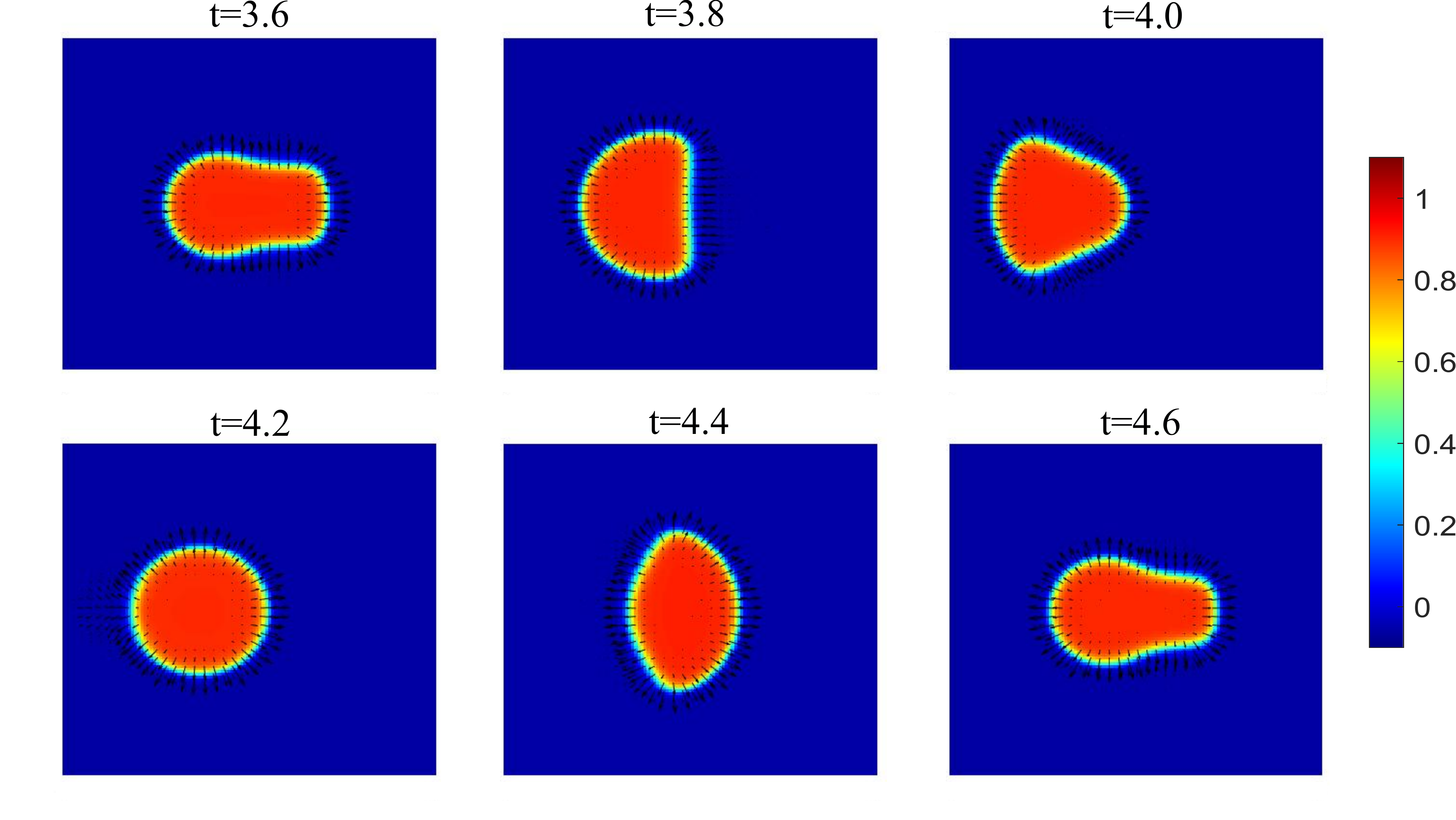}
\caption{By taking non-random initial conditions ($\rho$ circular and ${\bf P}=(1,0)$ on the interior of the cell), the model exhibits a two-wave state: at $t=3.8$ two waves of protrusion begin moving around the interface from the right hand side in opposite directions. At $t=4.0$ they collide on the left hand side of the cell, and at $t=4.4$ two waves similarly travel from left to right. Here we take $\ve = .03$ and $\beta = 130$. Color indicates value of $\rho$ and arrows represent the vector field ${\bf P}$.}
\label{fig:2wave}
\end{figure}

Rotating cells were also previously observed in the more sophisticated model of \cite{reeves2018rotating}, but it is perhaps surprising to be able to capture them in the current model without heterogeneity of myosin motors or adhesion dynamics.  We note however that more complex rotating modes, such as two-wave rotating states, were exhibited in \cite{reeves2018rotating}. {By carefully taking symmetric initial conditions, we were able to observe such two-wave rotating states, see Figure }\ref{fig:2wave}. {However, they are unstable, as a perturbation of initial conditions resulted in single-wave rotating states.} This suggests that heterogeneity of myosin/adhesion is required to stabilize such complex states.

Finally, our analysis showed that distance traveled (and thus cell velocity) is maximized for a fixed value of $\ve$, over a range of values of $\beta$. {Recalling original parameters, this suggests that a ratio of cell size to surface tension plays a more dominant role in controlling rotating cell speed than actin polymerization and adhesion site formation alone. Previous experiments on {\em Caenorhabditis elegans} sperm cells have found that membrane tension can optimize (traveling) cell speed \cite{batchelder2011membrane}: heuristically, if membrane tension is too small, polymerization occurs in many directions, impeding motion, whereas if tension is larger, polymerization can be focused to one direction.} 

We conjecture that by reincorporating heterogeneous myosin motor effects we can stabilize persistent motion over a wider range of parameter ranges so as to also obtain an optimal velocity occuring during persistent motion. 

{In order to relate simulations to the sharp interface limit equation, we were restricted to specific scalings related to $\ve$. 
	In future work it would be interesting to consider more general physical parameters in this simple model. Relaxing these restrictions would no longer relate the model to the sharp interface limit, though it would certainly allow for a wider range of quantitative biological comparisons.}

Finally, more rigorous bifurcation analysis between motility modes would be desirable: we anticipate existence of a Hopf bifurcation occuring from immobile to rotating states, and saddle-node bifurcations occuring between immobile and traveling wave states.

All simulations were completed in MATLAB and run on The College of New Jersey's high performance cluster, ELSA (Electronic Laboratory for Science and Analysis). Codes used to generate data can be provided upon reasonable request to the authors.

\section{Acknowledgments}

Portions of this research were completed using the high performance computing cluster (ELSA) at the School of Science, The College of New Jersey. Funding of ELSA is provided in part by National Science Foundation OAC-1828163. MSM was additionally supported by a Support of Scholarly Activities Grant at The College of New Jersey.
The authors would like to thank Dr. Nicholas Battista for helpful discussions during the writing of this manuscript and {Dr. Mykhailo Potomkin for insightful discussions on the non-dimensionalization of the model.} The authors also thank the referees for their helpful feedback which greatly improved the exposition.  NB additionally would like to thank The College of New Jersey Mathematics \& Statistics Department for its continuing support of undergraduate research, helping to propel his career.

\noindent {\em Declarations of interest:} none

\appendix

\section{Non-dimensionalization of the phase-field equations}\label{appendix}
\noindent { Starting from
	\begin{align}
	\partial_t \rho &= D_\rho \Delta \rho - \tau_1^{-1}W'(\rho) -\alpha (\nabla \rho)\cdot {\bf P}\\
	\partial_t {\bf P} &= D_P \Delta {\bf P} -\tau_2^{-1} {\bf P} -\zeta \nabla \rho,
	\end{align}
	introduce non-dimensional variables $T$ and $X$ via the diffusive scaling $t = T \frac{R^2}{D_\rho}$, and $x = X R$, where $R$ is a characteristic length scale of the cell, say the cell radius. Then
	\begin{align}
	\partial_T \rho &=  \Delta_{X} \rho - \frac{R^2}{D_\rho \tau_1}W'(\rho) -\frac{\alpha R}{D_\rho}(\nabla_X \rho)\cdot {\bf P}\\
	\partial_T {\bf P} &= \frac{D_P}{D_\rho}\Delta_{X} {\bf P} -\frac{R^2}{D_\rho \tau_2}{\bf P}-\frac{\zeta R}{D_\rho} \nabla_X \rho,
	\end{align}
	where $\nabla_X$ and $\Delta_X$ are derivatives with respect to the scaled variable $X$. We drop $X$ notation for clarity.
}

{	
	Introduce the dimensionless parameters
	\begin{equation*}
	\ve:=\frac{\sqrt{D_\rho \tau_1}}{R},\;\;\tilde{\alpha} := \frac{\alpha R}{D_\rho},\;\;\tilde{\zeta}:= \frac{\zeta\tau_1}{R}.
	\end{equation*}
	%
	By rescaling ${\bf P}\to \tilde{\alpha} {\bf P}$, we then have
	\begin{align}
	\partial_T \rho &= \Delta \rho - \frac{1}{\ve^2}W'(\rho) -  \nabla\rho \cdot {\bf P} \\
	\partial_T {\bf P} &= \frac{D_P}{D_\rho} \Delta {\bf P} - \frac{\tau_1}{\tau_2}\frac{1}{\ve^2} {\bf P}-\tilde{\alpha} \tilde{\zeta} \frac{1}{\ve^2} \nabla \rho
	\end{align}
	To obtain the model of \cite{berlyand2017sharp} requires the assumptions
	\begin{equation}
	D_P \sim \ve D_\rho,\;\; \tau_1\sim \ve \tau_2,
	\end{equation}
	and to define
	\begin{equation}
	\beta := \frac{\tilde{\alpha}\tilde{\zeta}}{\ve^2} = \frac{\alpha\zeta R^2}{D_\rho^2},
	\end{equation}
	resulting in
	\begin{align}
	\partial_T \rho &= \Delta \rho - \frac{1}{\ve^2} W'(\rho) -  \nabla\rho \cdot {\bf P} \\
	\partial_T {\bf P} &= \ve \Delta {\bf P} - \frac{1}{\ve} {\bf P}-\beta \nabla \rho.
	\end{align}
	The scalings $D_P\ll D_\rho$ and $\tau_1\ll \tau_2$ are consistent with experimental values for keratocytes reported in \cite{Zie12}, see also Table \ref{tab:1}.
}

%
%
%

\vfill\newpage
\bibliographystyle{amsplain}
\bibliography{cellref}

\end{document}